\documentclass[twocolumn,nopreprintnumbers,nolongbibliography,aps,prb,10pt,showpacs]{revtex4-1}% PRL
\usepackage{bm}% bold math
\usepackage{hyperref}
\usepackage{url}
\usepackage{amsmath}
\usepackage{amssymb}
\usepackage{amsxtra}
\usepackage{amscd}
\usepackage{amsthm}
\usepackage{amsfonts}
\usepackage{eucal}
\usepackage{latexsym,amsthm}
\usepackage{pstricks}
\usepackage{color}
\usepackage{graphicx}% Include figure files

\newcommand{\nc}{\newcommand}
\nc{\on}{\operatorname}
\nc{\wt}{\widetilde}
\nc{\Wick}{{\mathbb :}}
\nc{\R}{{\mathbb R}}

\newcommand{\beq}{\begin{equation}}
\newcommand{\eeq}{\end{equation}}
\newcommand{\bmul}{\begin{multline}}
\newcommand{\emul}{{\end{multline}}}
\newcommand\beqa{\begin{eqnarray}}
\newcommand\eeqa{\end{eqnarray}}
\newcommand\bea{\begin{array}}
\newcommand\eea{\end{array}}
\newcommand\ba{\begin{array}}
\newcommand\ea{\end{array}}
\newcommand{\nn}{\nonumber}

\newcommand{\neqa}{\nonumber\end{eqnarray}}

\newcommand{\eq}[1]{Eq.(\ref{#1})}

\newcommand{\Eq}[1]{Eq.(\ref{#1})}

\newcommand{\ur}[1]{(\ref{#1})}

\renewcommand{\d}{\partial}

\nc{\CH}{{\mathcal H}}
\nc{\Db}{{\bar D}}
\nc\comment[1]{}

\nc{\CM}{{\mathcal M}}
\nc{\CN}{{\mathcal N}}

\newcommand{\re}{\relax{\rm I\kern-.18em R}}

\renewcommand{\Im}{{\mathrm {Im}}}

\nc{\meV}{{\mathrm{\,meV}}}
\nc{\cG}{{\mathcal G}}

\renewcommand{\)}{\right)}
\renewcommand{\(}{\left(}
\renewcommand{\bar}{\overline} 

\nc{\al}{{\alpha}}

\def\eps{{\epsilon}}
\def\cO{{\cal O}}

\def\sign{{\rm \, sign }}

\renewcommand\red{{}}
\begin{document}

\title{Nonlinear magnetization of graphene}
\author{Sergey Slizovskiy}
\email{on leave from PNPI;  S.Slizovskiy@lboro.ac.uk}
\author{
Joseph J. Betouras}
\email{J.Betouras@lboro.ac.uk}
\affiliation{Department of Physics,  Loughborough University,\\
Loughborough LE11 3TU, UK}
\keywords{graphene, magnetization, magnetic properties, monolayer graphene}
\pacs{75.70.Ak , 73.22.Pr}
\begin{abstract}
We compute the magnetization of graphene in a magnetic field, taking into account for generality the possibility of a 
mass gap. We concentrate on the physical regime where quantum oscillations are not observed due to the effect of the temperature or disorder
and show that
the magnetization exhibits 
non-linear behaviour as a function of the applied field, reflecting the strong non-analyticity of the two-dimensional effective action of Dirac
electrons. 
The necessary values of the magnetic field to observe this
non-linearity vary from  a few Teslas for very clean suspended samples to 20--30 Teslas for good samples on substrate. In the light of
these calculations, 
we discuss the effects of disorder and interactions as well as the experimental
conditions under which the predictions can be observed.
\end{abstract}
\maketitle

\section{Introduction}
The physics of graphene has attracted a huge amount of interest since its discovery \cite{Novoselov} due to its unique physical properties as
well as its potential technological importance. 
The electronic properties and its behavior in a strong magnetic field have been the focus of wide activity as recent reviews 
summarize \cite{Geim2009,Goerbig2011}.

The Dirac-like spectrum of clean graphene has been established, see e.g. Ref.[\onlinecite{PartoensPeeters}] and references therein.
Moreover, the partition function of two-dimensional massless systems are known to exhibit
strong non-analytic behaviour as a function of external fields \cite{Redlich84, Dunne}. This kind of behaviour can lead, for
example, to
non-Ohmian conductivity due to Schwinger pair production at certain conditions \cite{Schwinger08,Schwinger10} or signatures of
quantum criticality \cite{Schmalian}. 
In the case of conventional metals, the magnetism is a result of two contributions, coming from the spin (Pauli contribution) or the
orbitals (Landau diamagnetism).
In this work, we
examine in detail the orbital magnetization of graphene in a magnetic field, including for generality the possibility of a mass gap,  and show the appearance 
of a non-linear dependence on the applied field, as a consequence of non-quadratic field-dependence of the partition function.
 The Pauli magnetization of graphene is linear as a function of doping and much smaller than the orbital magnetization especially for low carrier densities (chemical potential close to 0) \cite{LatticeEffects11}. The possible nonlinearity due to the Pauli contribution, which is a known phenomenon, a consequence of the saturation of localized moments at higher fields, is discussed in the part of this work where the experimental verification is proposed.

The nonlinearity happens as a result of the failure of the linear response approximation, which is applicable when there is a mass scale, larger than the
applied perturbation. This scale controls the perturbative expansion. For the case of graphene this could be e.g. a mass gap, the temperature or
the impurity scattering rate. Since the magnetic energy scale in graphene is known to be exceptionally large due its linear
dispersion, even at moderate magnetic fields the magnetic energy scale becomes dominant, and therefore it violates the basic assumptions of the linear response regime.

The non-linearity survives even at relatively small magnetic field for sufficiently clean samples, therefore it is not
sufficient to compute only the zero field magnetic susceptibility.  Typically, the observed non-linear effect of strong magnetic field
is the magnetic oscillations, but these oscillations average out at elevated temperatures or due to impurities producing the linear
magnetization. Here we focus on the regime, where the magnetic oscillations average out, but, nevertheless, 
a smooth non-linearity remains. 
%\comment{ The conditions to observe the nonlinear magnetization are to large extent the same as the conditions to observe the
%integer quantum Hall effect.}
In the following sections we discuss the effects of temperature, mass gap, disorder, as well as interaction effects at a
phenomenological level and propose an experiment which can demonstrate these predictions.

\section{Clean graphene with possible mass gap.} 
At zero temperature, the grand canonical potential of clean graphene in a magnetic field {\red when Landau levels (LLs) are formed}, 
at zero 
chemical potential, with energy of the LLs $E_n=\sign(n) \sqrt{\alpha |n B|}$ where $\alpha = 2 \hbar e v_F^2$ and $n$ is an integer reads
\beq \label{expansion standard}
 \Omega_{vac} = g_s g_v \sqrt{\alpha} |B|^{3/2} C \left(\frac{\zeta(3/2)}{4 \pi} - 0.1654\, a \sqrt{|B| C} \)
%- 0.06 a^2 B C\right) 
\nn
\eeq
with $a=0.142$ nm being the distance between the carbon atoms
%$\al = 2 \hbar e v_F^2$ 
and the degeneracy factor $C=\frac{e}{2 \pi \hbar}$, $g_v=2$, $g_s=2$ accounts for
spin and valley degeneracy, $\zeta$ is the zeta function: $\frac{\zeta(3/2)}{4 \pi}=-\zeta(-1/2) \approx 0.2079$. 

The subleading terms are lattice corrections which are discussed elsewhere \cite{LatticeEffects11, SlizovskiyBetouras2}, the
numerical factor corresponds to nearest-neighbour tight-binding model calculation and describes orbital paramagnetic contributions due
to higher energy levels having non-Dirac dispersion. The corrections as a result of the deviation of the dispersion from the Dirac-like
appear either due to interaction effects -- which we discuss below -- or
due to lattice effects when $B \gg 100$ Tesla and are not discussed further here. 

The leading term, though, is non-analytic in the magnetic field and naturally leads to divergent diamagnetic
susceptibility at the Dirac point. Such non-analyticities are typical not only in the effective action of massless systems but 
at quantum critical points 
\cite{Belitz0,Belitz1, ChubukovMaslov, EfremovBetourasChubukov} or even in corrections to Fermi liquid theory
\cite{BetourasEfremovChubukov}.

When temperature-averaged over many Landau levels (LLs), this non-analytic contribution cancels out, as first shown by McClure
\cite{McClure56}.
% because then the temperature 
%plays the role similar to the mass. 
When the magnetic field overcomes the energy scale set up by the temperature (or impurity scattering), this non-linearity 
can be observed.  For generality and to compare energy scales, we take into account a mass gap $\Delta$ in the dispersion relation, which can be experimentally
created e.g. due to A-B sublattice asymmetry caused
by SiC substrate  or by regular deposition of impurities \cite{Impurity_Lattice,note1}. 
At zero magnetic field it was demonstrated that the pseudo-spin degree
of freedom (due to valleys) produces diamagnetic susceptibility which is \cite{Gusynin04, Nakamura, Koshino10}:  
\beq \label{KoshinoGap}
\chi(\eps) = -g_s g_v \frac{\alpha C}{12 |\Delta|} \theta(|\Delta|-\eps) 
%=- g_s g_v \frac{e^2 v_F^2}{6 \pi} \frac{1}{2 |\Delta|}
%\theta(|\Delta|-|\eps|) 
\eeq 
it is evident that the mass gap resolves the formal $\delta$-function singularity in the susceptibility, but when the magnetic
energy scale exceeds that of the gap, the non-analyticities of the free energy become again dominant.   

In a magnetic field $B$ and in the presence of the mass gap, the spectrum at low energies  becomes
\beq
E_{n\neq 0} = \sign(n) \sqrt{\alpha |n B| + \Delta^2}  \ \ ;  \ \ E_0 = (-1)^v \Delta
\eeq
where $v=0,1$ enumerates the two Dirac valleys.
%The other non-linear correction, coming from the last term in \eq{expansion standard} is substantially smaller. 
In the absence of the gap or when its size is smaller than the distance between the first LLs $\Delta < \sqrt{\alpha
|B|}$,
the linear expression \ur{KoshinoGap} is not applicable and, as we will show, the correct result leads to the non-linear
magnetization.

 The regularized
free energy reads (at chemical potential $\mu=0$, temperature $T=0$)
\beqa
 \label{ExactZeta}
&\Omega(\Delta,0,0)&= - g_s C |B| \(\sum_{n=0}^\infty+\sum_{n=1}^\infty\) \sqrt{\alpha n |B| + \Delta^2} \nn \, \\  
&&=^{\!\!\!\!\!reg} -
g_s C \al^{1/2} |B|^{3/2} 
\(-\(\frac{\Delta^2}{\alpha |B|}\)^{1/2} + \right.\nn  \\  
&& \left. + 2\, \zeta\(-\frac12,
\frac{\Delta^2}{\al |B|}\) + \frac43 \(\frac{\Delta^2}{\al |B|}\)^{3/2}  \)
\eeqa
\noindent where we have regularized and subtracted the $B=0$ expression, $\zeta$ is a generalized Hurwitz $\zeta$-function. It also
agrees with related formula in Ref.[\onlinecite{Gusynin04}]. The
above expression is exact in the limit of free electrons and has a strong-field
expansion valid for $\delta^2 \equiv \frac{\Delta^2}{\al |B|} \lesssim 1$ (when the gap is less or comparable
to the distance between the first LLs), leading to {\red the strong-field expansion of} the magnetization ($\mu=0$, $T=0$): 
\beqa \label{MagnetizNonlin}
 &M(\Delta,0,0)= g_s C \sqrt{\al |B|} \(3  \zeta(-1/2) + \delta + \frac{\zeta(1/2)}{2} \delta^2  \right. \nn \\  
& \left. + \frac{\zeta(3/2)}{8}
\delta^4 -\frac{3 \zeta(5/2)}{16}
\delta^6 + \cO(\delta^8) \)
\eeqa
where $\delta \equiv \sqrt{\frac{\Delta^2}{\al |B|}} = 27.5 \frac{\Delta({\mathrm{eV}})}{\sqrt{|B|({\mathrm T})}}$, for example, for
$\Delta=0.1$eV this 
formula works starting from approximately 7.5 Tesla, indicating the scale from which the non-linearities occurs.
%, as seen in
%Fig.
%\ref{fig:ZeroTMagnetization}. 

%\begin{figure} 
%\centerline{
%\includegraphics[scale=0.8]{ZeroTMagnetization.eps}}
%\vspace{-0.3 cm}
%\caption{\label{fig:ZeroTMagnetization} 
%Non-linear magnetization at zero temperature, $\Delta=0.1 eV$, small- and high-field expansions are compared.}
%\end{figure}

%When the chemical
%potential is between $k$-th and $k+1$-th LLs,  the contribution of $k + (k + 1)$ levels coming from the two
%valleys must be taken into account: for $\mu^2 \in \(\al B k+\Delta^2,  \al B (k+1) +\Delta^2 \) $ 
%\beqa \label{MagnetizNonlin1}
%M(\Delta,\mu) &=& M(\Delta,0) - g_s C \(\Delta - (2 k+1)\mu  \right. \nn \\ 
%&&  \left. +  \sum_{n=1}^k \frac{2 \al n B + \Delta^2}{\sqrt{\al n B +
%\Delta^2}}\) 
%\eeqa
%
At $T=0$ and as $\Delta\to 0$ the range of the values of $B$ where linear magnetization occurs decreases to zero. Therefore, the
nonlinearity of magnetization gets stronger with the reduction of the gap and temperature.
{
The susceptibility, following from \eq{ExactZeta} reads: 
\beqa
&\chi(\Delta,0,0)=\frac{g_s  C \sqrt \alpha}{2 \sqrt{|B|}} \left[ 3 \zeta
   \left(-\frac{1}{2},\frac{\Delta^2}{\al |B|} \right)- \right.\\ \nn
& \left.- 2 \frac{\Delta^2}{\al |B|} \zeta
   \left(\frac{1}{2},\frac{\Delta^2}{\al |B|}\right)-\frac{\Delta^4}{\al^2 |B|^2}\zeta
   \left(\frac{3}{2},\frac{\Delta^2}{\al |B|}\right)\right]
\eeqa
}  
  
%At non-zero temperature we can use \cite{McClure56}:
%\beq \label{FiniteT formula}
 %M(\Delta,\mu, T) = -\int d \eps \,f'(\eps-\mu) M(\Delta,\eps, T=0)
%\eeq
%with $f$ being a Fermi distribution.
%%$-f'(\eps) = \frac{e^{\frac{\eps}{k_B T}}}{k_B T \(e^{\frac{\eps}{k_B T}}+1\right)^2}$. 
At finite temperature and chemical potential, after performing the same subtraction as for the regularization of
$\Omega(\Delta,\mu=0,T=0)$,
and additionally subtracting $\mu N(\mu=0,T=0)$ which is independent of $B$ we obtain \cite{comment2}:
\beqa \label{OmegaFull}
&\Omega&(\Delta,\mu, T) = \Omega(\Delta,0,0) - \\ \nn
&&-  g_s C B k T \left[\log\(1+\exp\(\frac{-\Delta+\mu}{k_B T}\)\) + \right. \\ \nn
&&\left. \!\!\!\!+2 \sum_{n=1}^{\infty}
\log\(1+\exp\(\frac{-E_n+\mu}{k_B T}\)\) 
\right] - \left[\mu\to -\mu\right]
\eeqa

\begin{figure}[t]  %%%% FIGURE 2
\begin{center}
\includegraphics[scale=0.8]{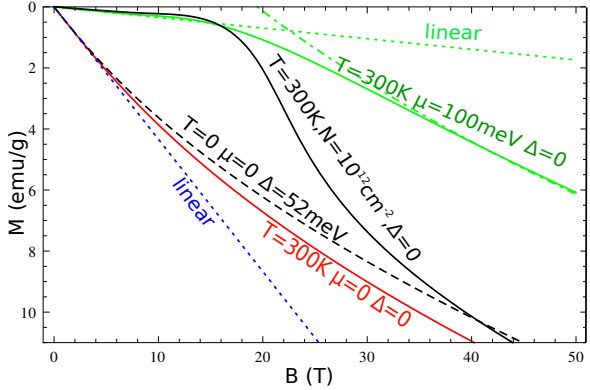}
\end{center}
\vspace{-0.6 cm}
\caption{\label{fig:TvsDelta} 
The magnetization at T=300 K, $\mu=0$ and at gap values $\Delta=0$ and 52 meV is compared to the linear behaviour. 
In addition, the magnetization at fixed chemical
potential 100 meV and at fixed electron concentration $10^{12} {\rm cm}^{-2}$ at T=300 K is plotted.
Low and high field asymptotic are shown.}
\end{figure}

At zero chemical potential the temperature plays a role similar to the gap. It is instructive to note that if we compare McClure's
result for
magnetic susceptibility at finite temperature: 
\beq
\nonumber
\chi = -\frac{g_s g_v C}{24} \frac{\al}{k_BT} {\rm sech}^2 \frac{\mu}{2 k_B T}
\eeq
\noindent with the one
for the magnetization with gap but at zero temperature, we 
see that at $2 k_B T = \Delta$ the linear magnetization is the same (with $\mu=0$), while the non-linear parts are different as is shown in Fig.
\ref{fig:TvsDelta}.  

For non-zero chemical potential and temperature there are two regimes, the well known low-field
regime and the high-field regime which sets in when $E_1=\sqrt{\alpha |B|} \gtrsim 2 \mu$.
When the chemical potential exceeds the temperature scale, the magnetization rapidly decreases as expected, but it starts to grow when
the
separation between the LLs becomes comparable to the chemical potential, this behaviour is shown by the
upper curve in Fig.\ref{fig:TvsDelta}. 
At such fields the zeroth LL gives the leading constant paramagnetic contribution to 
the magnetization, but the same non-linear vacuum energy contribution remains: for $\sqrt{\alpha |B|} \gg 2 (\mu+\Delta)
\gg k_B T $
\beq  \label{linear asymptotics}
 M(\Delta,\mu,T) \approx g_s C \, (\mu-\Delta) + M(\Delta,0,0) 
\eeq
this asymptotic regime is shown by the upper green dot-dashed line in Fig. \ref{fig:TvsDelta}.  When the temperature decreases we
get
de Haas - van Alphen oscillations for 
non-zero chemical potential, as expected \cite{Gusynin04}. The non-linearity we discuss can be alternatively interpreted as being
connected to this
dHvA oscillations, as its remnant behaviour. 

When the number of particles is fixed, instead of the chemical potential (both situations are
experimentally realizable in cases of graphene on substrate or suspended flakes) then $\mu$ is expressed through the relation
$N=-\frac{\d\Omega}{\d \mu}$. At small temperature, the 
De Haas van Alphen 
oscillations are observed. At high fields the chemical potential inevitably tends to zero since almost all the electrons (or holes if
$\mu<0$)  can
be hosted by the zero LL. This leads to the $|B|^{1/2}$ asymptotic behaviour of magnetization, corresponding to \Eq{ExactZeta}, shown by the black solid curve in Fig. \ref{fig:TvsDelta}. 

For completeness, we briefly comment on the case of bilayer graphene where
the zero-field susceptibility \cite{Safran,KoshinoMultilayer} diverges logarithmically with the Fermi energy $\eps_F\to 0$ and
the divergence is cut by the greatest of trigonal warping scale $\eps_{trig}$ and $\eps_F$. When we increase
the magnetic field, the magnetic energy scale $\sqrt{\al |B|}$ eventually becomes the greatest, thus leading to weak logarithmic non-linearity of the magnetization.  
The asymptotic form of the 
magnetic thermodynamic potential is \cite{Volovik12}: $\Omega = \frac{g_s g_v}{8 \pi} \frac{e^2
v_F^2}{\gamma_1} \log(\gamma_1/(\sqrt{\alpha |B|})) B^2 $ where $\gamma_1 \approx 0.4\,
{\mathrm eV}$ is the interlayer hopping energy and the magnetic scale $\sqrt{\al |B|}$ is assumed to be larger than the trigonal
warping energy $\eps_{trig}$ and $\eps_F$, otherwise one replaces $\sqrt{\al |B|} \to \eps_F$. Then one gets
\beq 
M \approx - |B| \frac{g_s g_v e^2 v_F^2}{8 \pi \gamma_1} \log\(\frac{\gamma_1^2}{|B| \al} \)
\eeq 
At even larger magnetic fields $B \gtrsim 100$ T, the magnetic energy becomes
larger than the interlayer hopping $\gamma_1$, thus effectively reducing the bilayer to two monolayers. 
Since the non-linearity of magnetization of graphene bilayer is significantly weaker than for the monolayer, we expect that the
impurities would make this effect hard to observe. So, we concentrate on the monolayer in what follows.  
The N-layered graphene was shown to have $[N/2]$ bilayer bands and $N \mod 2$ monolayer bands \cite{KoshinoMultilayer}.   

\section{Effect of impurities}
Besides the mass gap and temperature, the non-linearity of the magnetization is influenced by impurities and interactions.
We consider first the short-range scattering impurities and then the effect of charge inhomogeneities: the electron and hole puddles.  
\subsection{Short-range impurities}
Consider for simplicity the short-range impurities with momentum independent scattering. 
It is sufficient to adopt the self-consistent
Born approximation (SCBA)  \cite{ShonAndo98,Koshino07}. 
The treatment of the  vacancy-type impurities \cite{CastroNeto06}, or the commonly used phenomenological 
Lorentzian broadening leads to similar conclusions qualitatively, as we have checked.

%Here we prefer to estimate the effect of impurities and interactions from the experimental data to be more realistic.

%Instead of considering only the Dirac part of dispersion and using 
%a cut-off, we consider the full spectrum of nearest-neighbour hopping model. Then the expression for
%magnetic susceptibility gets an extra contribution, corresponding to Landau-like magnetism of the parabolic
%bottom of the dispersion.    
%
In SCBA \cite{Koshino07}, the self-energy reads
\beq \label{Born}
\Sigma(\eps) = \frac{W \alpha |B|}{2} \sum_n \frac{g(\eps_n)}{\eps - \eps_n - \Sigma(\eps)}
\eeq 
where $\eps_n$ is a full spectrum and $g(\eps)$ is some smooth cut-off function. This is valid for small strength of 
disorder $W=\frac{n_i u_i^2}{4 \pi v_F^2} \ll 1$, where $n_i$ is the impurity concentration and $u_i$ a measure of the strength of the random on-site impurity interaction.
For discrete LL spectrum and for small $W$ we obtain the solution iteratively.
The density of states is
\beq
\rho(\eps) = - \frac{g_v g_s}{2 \pi^2 \hbar^2 v_F^2 W} \Im\, \Sigma(\eps+ i 0)
\eeq

For the strong-field regime  the single-level approximation of the level width works well, as it has been checked numerically.
Solving the
\eq{Born} with a single energy level we obtain two roots and the relevant solution is: 
$\Sigma(\eps)=\frac12 \(\eps-\eps_n - \sign(\eps-\eps_n) \sqrt{(\eps-\eps_n)^2 - 2 \al W B}\) $, 
\comment{ $}
therefore the density of states can
be approximated
by a semi-circle form
\beq
\rho(\eps) = \frac{g_v g_s C}{\pi \al W} \sqrt{2 \al W |B| - \eps^2}\, {\mathrm\theta}(2 \al W |B|-\eps^2)
\eeq
where $\theta$ is a step-function and $\eps$ means the deviation from the LL of the clean system. In the single-level approximation,
we neglect the shift of the level centre due to the real part of the self-energy. 
Similar form for the density of states is supported by more elaborate computations, see e.g. \cite{Dora}.
Note that the level width scales as $\sqrt{|B|}$.
The above approximation can be applied for levels $\eps_n$ up to $n \lesssim \frac{1}{8 W}$, for higher levels their overlap would
become essential. Since at medium or strong fields the higher levels are typically far from the Fermi surface, we may use
the same level-broadening for all the
levels, since this does not alter the result for the magnetization. 
%Note that for the zero LL the width may be substantially different, and this depends on the model for impurities, e.g.
%for a long-range model
%the width is estimated to be $\sqrt 2$ times bigger \cite{Koshino07}. 

Under the assumption that 
all the levels are broadened with the same profile  $\rho(\eps)$, the partition function can be
computed by integrating \eq{OmegaFull} with the broadened chemical potential:
\beqa \label{impurityOmega}
 \Omega_{imp} = \int d\eps\, \frac{\rho(\eps-\mu)}{\int \rho^2(\eps') d\eps'} \, \Omega(\Delta,\eps,T)  
\eeqa
We note that this equation automatically inherits the regularization from $\Omega(\Delta,\eps,T)$.
For the magnetization one gets a similar formula, but with an extra contribution coming from the $B$-derivative of the broadening profile.
At high fields this contribution is paramagnetic due to broadening of the zero LL.
As mentioned above, there is no
significant dependence on the actual width of higher LLs. 
%The results do not change qualitatively if we use other forms of LL
%broadening (e.g. Lorentzian).  
The 
effect of temperature may be alternatively taken into account, by convolving the impurity-broadening with the derivative of the 
Fermi function with respect to energy $f'$ \cite{comment1}. 

\begin{figure}[t] 
\begin{center}
\includegraphics[scale=0.8]{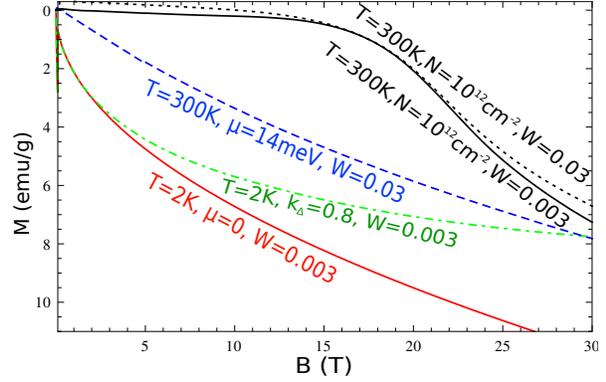}
\end{center}
\vspace{-0.7 cm}
\caption{\label{fig:Mu=14} 
Non-linear magnetization with short-range impurities for fixed $\mu$ or for fixed $N$ . 
Green dashed curve takes into account the linear in $B$ splitting of zero LL due to interactions, while the
green dot-dashed curve is for $\sim \sqrt{B}$ splitting of zero LL -- see Section \ref{sec:Interactions}. 
%Black solid curve is a magnetization
%for fixed electron concentration
}
\end{figure}

%The level width grows as $\sqrt{B}$ at high magnetic field which is of particular interest in this work. 
The typical line 
width of the high-quality sample  \cite{LLSpectroscopy06}, \cite{InteractionSpectroscopy10} is estimated to be
$\delta \approx 3 \sqrt{B({\rm T})} \meV$, so $W\approx 0.003$ and the 
Fermi energy is $\mu \approx 14 \meV$. Such high-quality sample is close to the ideal case and exhibits strongly
non-linear $\sim{\sqrt{|B|}}$ magnetization already at 1 Tesla, {\red as illustrated in} Fig. \ref{fig:Mu=14}. 

From this subsection we conclude that low concentrations of short-range impurities do not significantly alter the magnetization.
For larger concentrations of impurities the above analysis is not applicable.

\subsection{Charge inhomogeneities} 
It was shown experimentally, that for many graphene samples the charge inhomogeneities play
the dominant role \cite{DOS in Graphene}.  The effect of charge inhomogeneities can be modelled
as long-range smooth variation of carrier density $N$ (electron and hole puddles), 
and the range may be assumed to be
comparable to the size of cyclotron orbits. A simple model proposed in Ref. [\onlinecite{DOS in Graphene}] 
fits well the experiment. It assumes Gaussian variation of $N$ with standard deviation 
of order $\delta N \approx 4 \cdot 10^{11} \mathrm{cm}^{-2}$.  Since $N \sim \mu^2$, the variation $\delta N$ leads to large
$\delta \mu$ close to the tip of the Dirac cone and results in the large broadening of the
zeroth LL.   This contrasts with the equal profile broadening, coming from the scattering
on short-range impurities. Apart from variation of the charge density, the charged  impurities cause a usual broadening 
of the levels. We use the simple
model of constant Lorentzian broadening, independent of the magnetic field, as compared to short-ranged impurities, where we
got the $\sqrt{B}$ dependence. This does not
significantly change the result and is partly justified by computations in Ref. [\onlinecite{PeetersVshape}], where the $B$ dependence
was shown
to be more shallow than $\sqrt{B}$ due to the fact that the screening increases with $B$. 

\begin{figure}[t] 
\begin{center}
\includegraphics[scale=0.75]{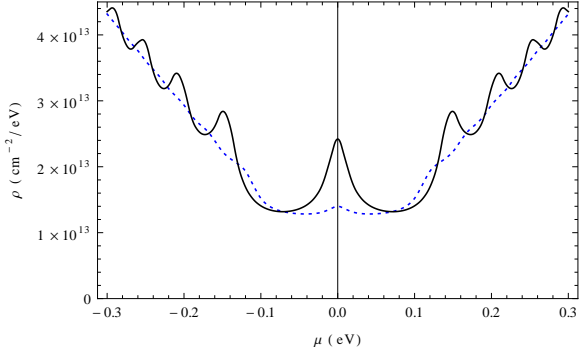}
\end{center}
\vspace{-0.7 cm}
\caption{\label{fig:DOS} Density of states for 16 T (solid) and 10 T (dotted) as a function of chemical potential of neutral spot with
gaussian carrier density fluctuation $\delta N \approx 4 \cdot 10^{11} \mathrm{cm}^{-2}$ and Lorentzian broadening of levels with
half-width 
%$\eps = 4 \sqrt{B(\mathrm{T})}
$\eps = 15 \meV$.
}
\end{figure}

To perform calculations, we broaden the levels with Lorentz profile, getting the density of states $\rho_\eps(E)$ then integrate it to
find $N(E_F)$ and it's inverse $E_F(N)$  and then
contract with Gaussian density fluctuation profile:
\beq
 P(N, \bar N, \delta N)=\frac{1}{\sqrt{2 \pi} \delta N} \exp\(-\frac{(N-\bar N)^2}{2 \delta N^2}\)
\eeq 
\noindent For example, the density of states is given by
\beq
 \rho(\bar N) = \int dN \, \rho_\eps(E_F(N)) \, P(N, \bar N,\delta N)    
\eeq  
The consideration of graphene with fixed total number of electrons is equivalent to the situation of graphene on substrate and 
the particle number being proportional to gate voltage. For suspended graphene (or exfoliated flake) one may imagine 
a situation of the fixed
local chemical potential at spots where it touches a contact. Assuming that these spots are far from the charged impurity (neutral
spot), the local electron density at such spots would
coincide with the average value $\bar N$, so we may use the same as above function $N(E_F)$ to convert $\rho(\bar N)$ to
$\rho(N(\mu))$. 
The resulting density of
states (DOS) as a function of the chemical potential is plotted in the figure \ref{fig:DOS} for magnetic fields 10 and
16 Tesla, $\delta N \approx 4 \cdot 10^{11} \mathrm{cm}^{-2}$ and Lorentzian broadening with half-width 
%$\eps = 4 \sqrt{B(\mathrm{T})}
$\eps = 15 \meV$. The same figure but plotted against the electron density $N$ (or gate voltage) can be found in Ref.
[\onlinecite{DOS in Graphene}].  
For high temperatures, the level broadening is dominated by temperature, and for low temperatures -- by impurities.

For low
temperatures and Lorentz impurity broadening $\eps$ the \eq{OmegaFull} for the grand canonical potential, combined with
\eq{impurityOmega} gives at $T=0$:
\beqa
\label{OmegaFullLorentz}
&\Omega(\Delta, \mu,\eps)= \Omega(\Delta,0,0) + \\
& \nn+g_s C B \left[F(\mu+\Delta)+ 2 \sum_{n=1}^{\infty} F(\mu+E_n)\right] + [\mu
\to -\mu] 
\eeqa 
where $E_n=\sqrt{\alpha n |B| + \Delta^2}$,
\beqa
&F(e) = \frac{e}{\pi} (\arctan(\max(\Lambda,e)/\eps)-\arctan(\max(-\Lambda,e)/\eps)) \nn 
\\ 
& + \frac{\eps}{2 \pi} \log
\frac{\max(e,-\Lambda)^2+\eps^2}{\max(e,\Lambda)^2+\eps^2} 
\eeqa
and $\Lambda$ is a large cut-off for Lorentzian broadening. 

To see the effect of charge puddles on magnetization, we compute the magnetization as a function of electron density $N$ and then
convolve with Gaussian profile:
\beq
M(\bar N,\delta N, T) = \frac{1}{\sqrt{2 \pi} \delta N} \int dN M(N,T) e^{-\frac{(\bar N - N)^2}{2 (\delta N)^2}}
\eeq  
this prescription follows from summation over separate puddles, $\bar N$ denotes the average electron density.   
 
\begin{figure}[t] 
\begin{center}
\includegraphics[scale=0.77]{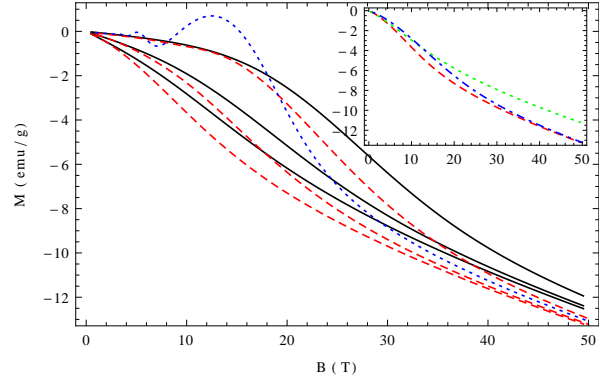}
\end{center}
\vspace{-0.7 cm}
\caption{\label{fig:puddlesn} 
Non-linear magnetization with charged impurities for fixed average electron density. Bottom to top: $\bar N=0 ;\ 5\cdot 10^{11} ;\ 
10^{12} ; 1.5\cdot 10^{12}\mathrm{cm}^{-2} $, temperature $T=300$ K (black solid) and $T=0$ (red dashed);  Lorentz broadening $\eps=15
\meV$, density fluctuation dispersion is $\delta N = 4\cdot 10^{11} \mathrm{cm}^{-2}$. A plot for smaller $\delta N = 4\cdot 10^{10}
\mathrm{cm}^{-2}$ with $\bar N=10^{12}\mathrm{cm}^{-2}$ and $T=0$ (blue dotted) is shown for comparison.\\
Inset: Dependence on the impurity strength. Red dashed curve is the same as the bottom one on the main plot: $\bar N = 0$, $T=0$,
$\eps=15 \meV$, $\delta N = 4\cdot 10^{11}
\mathrm{cm}^{-2}$;
green dotted: Lorentz broadening increased to $\eps=30 \meV$; blue dot-dashed: density fluctuation increased to $\delta N = 6\cdot
10^{11} \mathrm{cm}^{-2}$. 
}
\end{figure}

\begin{figure}[t] 
\begin{center}
\includegraphics[scale=0.77]{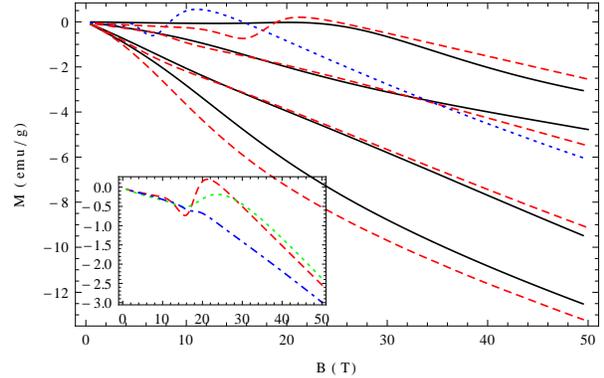}
\end{center}
\vspace{-0.7 cm}
\caption{\label{fig:puddlesmu} 
Non-linear magnetization charged impurities for fixed chemical potential of neutral spot. 
Bottom to top: $\mu=0; \ 50 ;\ 100; \ 150 \meV$  temperature $T=300$ K (black solid) and $T=0$ (red dashed);  Lorentz broadening is
$\eps=15 \meV$,  density
fluctuation dispersion is $\delta N = 4\cdot 10^{11} \mathrm{cm}^{-2}$. 
A plot for smaller $\delta N = 4\cdot 10^{10} \mathrm{cm}^{-2}$, $\mu = 100 \meV$, $T=0$ 
(blue dotted) is shown for comparison.\\
Inset: Dependence on the impurity strength. Red dashed curve is the same as on the main plot:
$\mu=150 \meV$, $T=0$, $\eps=15 \meV$, $\delta N = 4\cdot 10^{11}
\mathrm{cm}^{-2}$;
green dotted: Lorentz broadening increased to $\eps=30 \meV$; blue dot-dashed: density fluctuation increased to $\delta N = 6\cdot
10^{11} \mathrm{cm}^{-2}$. 
}
\end{figure}

The results for fixed average carrier number $\bar N$ and for fixed chemical potential $\mu$ of neutral spot are shown in Figs. 
\ref{fig:puddlesn}, \ref{fig:puddlesmu}.
From these plots we see that charge disorder observed in experiments plays an important role to smear the magnetic oscillations (the 
plot for lower charged disorder shows clear magnetic oscillations),
as does also the temperature, but the resulting magnetization is still non-linear.  The non-linearity of magnetization gets weaker
with increase of both types of disorder, but in a different way -- see the insets to Figs. \ref{fig:puddlesn}, \ref{fig:puddlesmu}.

%At high magnetic fields and low temperatures for fixed average $\mu$
%the averaged chemical potential is inevitably between 0 and the first LL, so, near the charge puddles the local chemical potential
%is either $\mu \approx 0$ or $\mu \approx  E_1=\sqrt{\alpha |B|}$ depending on the puddle charge.  For symmetric distribution of
%puddle charges (as is assumed in the plots), the high field asymptotics of magnetization is then given by \eq{linear
%asymptotics}
%with $\mu = \frac{0 + \sqrt{\alpha |B|}}{2}$.  

\section{Effect of interactions} \label{sec:Interactions}
It is worth noting that the most drastic effect of interactions is the splitting of the zero LL 
into two levels separated by a new gap $2 \tilde\Delta$ (different from the initial $\Delta$), when $\mu \approx 0$ and the zero LL is not
completely filled \cite{FirstSplitting,ZeroGap09,InteractionSpectroscopy10}. 
$\tilde\Delta$ grows with magnetic field. The precise form of its dependence on the magnetic field is an open question and depends on the sample. 
Generally, there are the following kinds of energy gaps as a consequence of the magnetic field:(i)  linear in $B$ dependence,  coming from the Zeeman spin splitting $\Delta_Z = 2 \mu_B B \approx 0.11 B({\mathrm T})
{\meV}$ and from the potential pseudospin splitting (from Kekule-type distortion of the lattice) $\Delta_{Kekule} \approx 0.2
B({\mathrm T}) \meV$, and (ii) interaction contributions, scaling as $\sqrt{B}$,  as explained e.g. in Ref.[\onlinecite{Goerbig2011}] and references therein. 

The experimental data in Ref.[\onlinecite{InteractionSpectroscopy10}] contain significant uncertainty, allowing for various
fits. For samples with high mobility $17000 \mathrm{cm^2/(V s)}$ the fit by $\sqrt{B}$ looks reasonable: $\tilde\Delta \approx
k_{1/2}\,\sqrt{|B|}$, with $k_{1/2} \approx 3 \meV/\sqrt{\mathrm{T}}$. For samples with
lower mobility  a fit linear in B is within error bars:
$\tilde\Delta \approx k_{linear}\,|B|$ with $k_{linear} \approx 0.8 \meV/{\mathrm T}$ at half-filling.
This is 7 times larger than the Zeeman splitting, indicating a different mechanism.

The splitting of the zeroth LL
leads to an extra paramagnetic contribution, as compared to \eq{OmegaFull} with $\tilde\Delta=0$, $\mu=0$.
The correction to thermodynamic potential is easily computed by replacing $\log 2 \to \frac12 \log\(1+e^{-\tilde\Delta/(k_B
T)}\)+\frac12 \log\(1+e^{+\tilde\Delta/(k_B T)}\)$. 
\comment{$}  
For the linear dependence
on $B$, defining $x=\frac{k_{linear}\, |B|}{k_B T}$  we have  
in the single-particle
approximation: 
\beqa
\delta M &=& C g_s k_B T \(2 \log\cosh\frac{x}{2} + x \tanh \frac{x}{2}\)
  \nn
\eeqa  
This is shown by the dot-dashed curve on Fig. \ref{fig:Mu=14}. 
The paramagnetic nature of this contribution can be understood as due to the reduction in energy of the filled half 
of the zeroth LL by interactions. The correction is quadratic for small $x$ and linear at large $x$. 

For $\sqrt{B}$ splitting, the correction of the magnetization is 
\beqa
\delta M &=& C g_s k_B T \(2 \log\cosh\frac{x}{2} + \frac12 x \tanh \frac{x}{2}\)
  \nn
\eeqa    
with $x=\frac{k_{1/2}\,\sqrt{|B|}}{k_B T}$.
%It starts to be non-linear when $B/T \gtrsim 1\, 

The different fittings of the splitting of the zeroth LL as a function of B, lead to different B-dependence of the magnetization,
which, in turn, can provide one more method to distinguish between the main possibilities. 
 We note that still there is no fully satisfactory theoretical explanation of the level-splitting effect
\cite{InteractionSpectroscopy10}, and its microscopic explanation is beyond the scope of this work.

\section{Experimental Proposal and discussion}

 The measurements of non-linear magnetization can be used to study various properties of graphene
sample: magnetization is sensitive to
the number of carriers,  mass gap and disorder as well as number of layers. In particular, one can extract the magnetic field dependence of the
interaction-induced splitting of the zero LL.  
One possible way to measure the nonlinear magnetization is to measure the
magnetization of a suspended graphene flake with  scanning SQUID microscopy.  Alternatively,
one may go to higher magnetic fields (up to 50 Tesla) with larger amount of lower-quality graphene samples, e.g. the graphene
laminate, used in Ref.[\onlinecite{MagnetizationGeim10}].  At high fields the cyclotron orbits would fit better in the small-sized
flakes 
and one can neglect the boundary effects.  In real samples there will be a significant non-linear part of the magnetization, 
coming from localized impurity spins (Pauli contribution),  but this effect can be fitted at lower magnetic fields or fields parallel
to the surface, and consistently subtracted \cite{Nair}. 
Moreover, at magnetic fields greater or of the order of 10 Tesla and at low temperatures $T<4$ K, it is
expected that all the localized moments will come to saturation, therefore the detected nonlinearity will be a consequence of the
orbital contribution.

In conclusion, the two-dimensional nature and the linear spectrum of graphene are the necessary conditions to observe
non-linear magnetization at accessible magnetic fields. The underlying reason is the breakdown of the linear response theory due to the fact that the magnetic energy is the dominant in the system. The linear dispersion relation gives relatively
large distances between the LLs and the two-dimensionality leads to absence of $k_z$ dispersion of LLs.  We
have found that even at room temperature and with moderate concentration of impurities  the non-linearity should be revealed at
about 10-20 Tesla,
while with very clean suspended samples at low temperatures a lower value of magnetic field is sufficient.   There are two types of
non-linear behaviour that are present:  near half-filling, the magnetization scales as $\sqrt B$ at higher fields, due to
non-analyticity of
the effective action; then at higher values of the chemical potential, the magnetization is small and linear at small fields, while it increases the
slope after the first LL crosses the Fermi energy, and at even higher magnetic fields one can also observe the $\sqrt B$ behaviour.

We acknowledge inspiring discussions with Feo Kusmartsev, Marat Gaifullin, Roberto Soldati and discussions of experimental results
with Irina
Grigorieva and Paulina
Plochocka. This work was supported by the Engineering and Physical Sciences Research Council under EP/H049797/1. 
\vspace{-0.7cm}

\setlength{\bibsep}{1pt}

%\setlength{\bibsep}{1pt}
%\bibliographystyle{prsty}
%  \bibliography{MyReferences}

\begin{thebibliography}{10}
\bibitem{Novoselov}
 K.~S. Novoselov, A.~K. Geim, S.~V. Morozov, D. Jiang, Y. Zhang, S.~V. Dubonos, I.~V. Grigorieva, and A.~A. Firsov,  Science, {\bf 306}, 666 (2004)

\bibitem{Geim2009}
A.~H. {Castro Neto}, F. Guinea, N.~M. Peres, K.~S. Novoselov, and A.~K. Geim, Rev .Mod. Phys. {\bf 81},  109
  (2009).

\bibitem{Goerbig2011}
M.~O. {Goerbig}, Rev. Mod. Phys. {\bf 83},  1193  (2011).

\bibitem{PartoensPeeters}
B. Partoens and F.~M. Peeters, Phys. Rev. B {\bf 74}, 075404 (2006).

\bibitem{Redlich84}
A.~N. {Redlich}, \prd {\bf 29},  2366  (1984).

\bibitem{Dunne}
D. {Cangemi} and G. {Dunne}, Ann. Phys. {\bf 249},  582  (1996).

\bibitem{Schwinger08}
D. {Allor}, T.~D. {Cohen}, and D.~A. {McGady}, \prd {\bf 78},  096009  (2008).

\bibitem{Schwinger10}
B. {D{\'o}ra} and R. {Moessner}, \prb {\bf 81},  165431  (2010).

\bibitem{Schmalian}
M. Mueller, J. Schmalian, L. Fritz, Phys.Rev.Lett. {\bf 103}, 025301 (2009)  

\bibitem{LatticeEffects11}
G. G\'omez-Santos and T. Stauber, Phys. Rev. Lett. {\bf 106},  045504  (2011).

\bibitem{SlizovskiyBetouras2}
S. {Slizovskiy} and J.~J. {Betouras}, in preparation  .

\bibitem{Belitz0}
D. Belitz, T.~R. Kirkpatrick, and T. Vojta, Phys. Rev. B {\bf 55},  9452
  (1997).

\bibitem{Belitz1}
D. Belitz, T.~R. Kirkpatrick, A.~J. Millis, and T. Vojta, Phys. Rev. B {\bf
  58},  14155  (1998).

\bibitem{ChubukovMaslov}
A.~V. Chubukov and D.~L. Maslov, Phys. Rev. B {\bf 68},  155113  (2003).

\bibitem{EfremovBetourasChubukov}
D.~V. {Efremov}, J.~J. {Betouras}, and A. {Chubukov}, \prb {\bf 77},  220401
  (2008).

\bibitem{BetourasEfremovChubukov}
J. Betouras, D. Efremov, and A. Chubukov, Phys. Rev. B {\bf 72},  115112
  (2005).

\bibitem{McClure56}
J.~W. {McClure}, Physical Review {\bf 104},  666  (1956).

\bibitem{Epitaxial1}
S.~Y. {Zhou}, G.-H. Gweon, A.~V. Fedorov, P.~N. First, W.~A. de Heer, D.-H. Lee, F. Guinea, A.~H. Castro Neto, and A. Lanzara, Nature Materials {\bf 6},  916  (2007).

\bibitem{Epitaxial2}
S.~Y. {Zhou}, D.~A. Siegel, A.~V. Feodorov, F. El Gabaly, A.~K. Schmid, A.~H. Castro Neto, D.-H. Lee, and A. Lanzara, Nature Materials {\bf 7},  259  (2008).

\bibitem{Impurity_Lattice}
H. Sahin and S. Ciraci, Phys. Rev. B {\bf 84}, 035452 (2011).

\bibitem{note1}
A band gap of order 0.1 eV has been
observed in graphene epitaxially grown on SiC substrate
\cite{Epitaxial1,Epitaxial2}, but a word of caution is needed here, since substrate may induce high impurity concentrations.

\bibitem{Gusynin04}
S.~G. Sharapov, V.~P. Gusynin, H. Beck, \prb {\bf 69}, 075104 (2004)

\bibitem{Nakamura}
M.Nakamura, \prb{\bf 76}, 113301 (2007)

\bibitem{Koshino10}
M. {Koshino} and T. {Ando}, \prb {\bf 81},  195431  (2010).

\bibitem{comment2}
In the context of dynamical flavor symmetry breaking in 2+1 relativistic field theory models, 
related work can be found in V.~P. Gusynin, V.~A. Miransky, I.~A. Shovkovy, \prd {\bf 52}, 4718 (1995);
K.~G. Klimenko, Z.~Phys.~C {\bf 54}, 323 (1992);  K.G. Klimenko, Theor. Math. Phys. {\bf 90}, 1, (1992).

\bibitem{KoshinoMultilayer}
M. {Koshino} and T. {Ando}, \prb {\bf 76},  085425  (2007).

\bibitem{Safran} 
S. A. Safran,  \prb {\bf 30}, 421 (1984).

\bibitem{Volovik12}
M.~I. {Katsnelson} and G.~E. {Volovik}, arXiv: 1203.1578.

\bibitem{Koshino07}
M. {Koshino} and T. {Ando}, \prb {\bf 75},  235333  (2007).

\bibitem{Dora} B.~D{\'o}ra, Low 
Temp. Phys. {\bf 34}, 801 (2008)

\bibitem{ShonAndo98}
N. {Shon} and T. {Ando}, Journal of the Physical Society of Japan {\bf 67},
  2421  (1998).

\bibitem{CastroNeto06}
N.~M.~R. {Peres}, F. {Guinea}, and A.~H. {Castro Neto}, \prb {\bf 73},  125411
  (2006).

\bibitem{comment1}
 \Eq{impurityOmega} can be straighforwardly generalized to the case when the zero LL has a different broadening
profile $\rho_0$
by adding $\int_{-\infty}^\mu d\eps\, \eps \,(\rho_0(\eps)-\rho(\eps))$ where the profiles need to be convoluted 
with the derivative of the Fermi function $f'$ to account for temperature.  

\bibitem{LLSpectroscopy06}
M.~L. Sadowski,  G. Martinez, M. Potemski, C. Berger, and W.~A. de Heer, Phys. Rev. Lett. {\bf 97},  266405  (2006).

\bibitem{PeetersVshape} C. H. Yang, F. M. Peeters, and W. Xu 
%Density of States and magneto-optical conductivity of graphene in a
%perpendicular magnetic field, 
Phys. Rev. {\bf B 82}, 205428 (2010). 

\bibitem{FirstSplitting} Y.~Zhang, Z. Jiang, J.~P. Small, M.~S. Purewal, Y.-W. Tan, M. Fazlollahi, J.~D. Chudow, J.~A. Jaszczak, H.~L. Stormer, and P. Kim, Phys. Rev. Lett. {\bf 96},  136806  (2006).

\bibitem{InteractionSpectroscopy10}
E.~A. Henriksen,  P. Cadden-Zimansky, Z. Jiang, Z.~Q. Li, L.-C. Tung, M.~E. Schwartz, M. Takita, Y.-J. Wang, P. Kim, and H.~L. Stormer, Phys. Rev. Lett. {\bf 104},  067404  (2010).

\bibitem{DOS in Graphene}
 L.~A.Ponomarenko et al.  Phys. Rev. Lett. {\bf 105}, 136801 (2010)

\bibitem{ZeroGap09}
A.~J.~M. {Giesbers}, L~ A. Ponomarenko, K.~S. Novoselov, A.~K. Geim, M.~I. Katsnelson, J.~C. Maan, and U. Zeitler, \prb {\bf 80},  201403  (2009).

\bibitem{MagnetizationGeim10}
M. Sepioni, R.~R. Nair, S. Rablen, J. Narayanan, F. Tuna, R. Winpenny, A.~K. Geim, and I.~V. Grigorieva, Phys. Rev. Lett. {\bf 105},  207205  (2010).

\bibitem{Nair}
R.~R. Nair, M. Sepioni, I-Ling Tsai, O. Lehtinen, J. Keinonen, A.~V. Krasheninnikov, T. Thomson, A.~K. Geim, and I.~V. Grigorieva, Nature Physics {\bf 8}, 199 (2012)

\end{thebibliography}
%\end{spacing}
\end{document}